\documentclass[runningheads]{llncs}
\usepackage{graphicx}
\usepackage[left=3.0cm, right=3.0cm, top=3.cm  ]{geometry}
\usepackage{graphicx}
\usepackage{amsfonts} 
\usepackage{amsmath}
\usepackage{dsfont}
\usepackage{mathtools}
\usepackage[linesnumbered,ruled,vlined]{algorithm2e}
\usepackage{subcaption}

\usepackage{caption}

\usepackage{hyperref}
\usepackage[T1]{fontenc}
\usepackage[rgb]{xcolor}
\usepackage{smartdiagram}

\begin{document}

\title{Impact of delay classes on the data structure in IOTA}

\author{Andreas Penzkofer, Olivia Saa and Daria Dziubałtowska}
\institute{IOTA Foundation, Berlin, Germany
}

\maketitle
\thispagestyle{plain}
\pagestyle{plain}

\begin{abstract}

In distributed ledger technologies (DLTs) with a directed acyclic graph (DAG) data structure, a message-issuing node can decide where to append that message and, consequently, how to grow the DAG. 
This DAG data structure can typically be decomposed into two pools of messages: referenced messages and unreferenced messages (tips). The selection of the parent messages to which a node appends the messages it issues, depends on which messages it considers as tips. 
However, the exact time that a message enters the tip pool of a node depends on the delay of that message. 
In previous works, it was considered that messages have the same or similar delay; however, this generally may not be the case. 
We introduce the concept of classes of delays, where messages belonging to a certain class have a specific delay, and where these classes coexist in the DAG.
We provide a general model that predicts the tip pool size for any finite number of different classes. 

This categorisation and model is applied to the first iteration of the IOTA 2.0 protocol (a.k.a. Coordicide), where two distinct classes, namely value and data messages, coexist. 
We show that the tip pool size depends strongly on the dominating class that is present. Finally, we provide a methodology for controlling the tip pool size by dynamically adjusting the number of references a message creates.

\end{abstract}

\section{Introduction}

\begin{figure}[b!]
\vspace{-0.3cm}
\center
\begin{minipage}{1.\textwidth}
    \center
    \includegraphics[width=.8\textwidth,trim={0 0 0 0},clip]{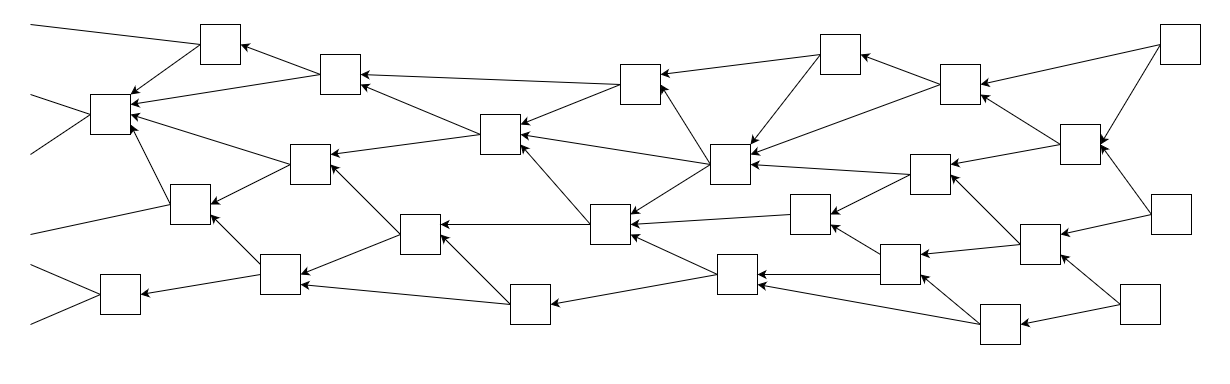}
    
    \small a) One delay class of message.
    
    \includegraphics[width=.8\textwidth,trim={0 0 0 0},clip]{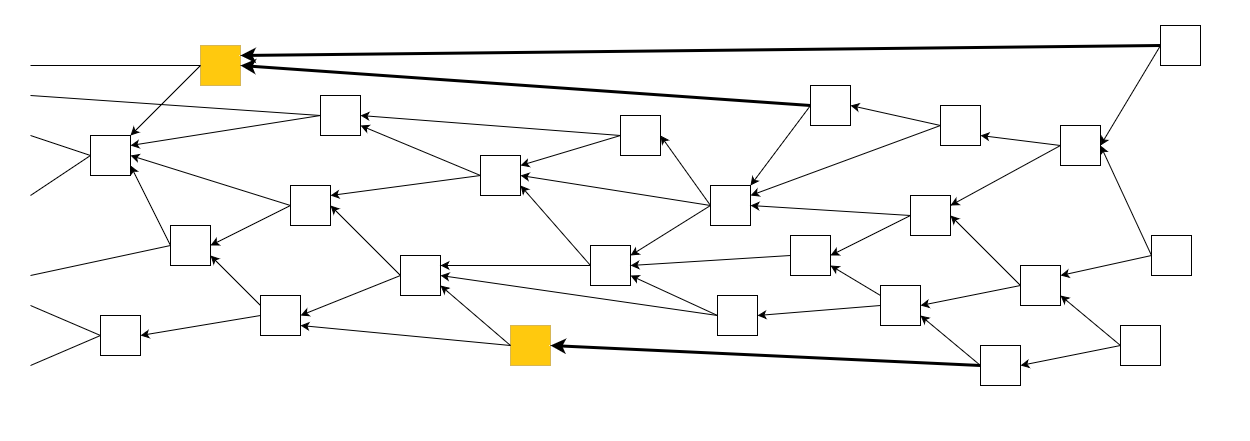}
    
    b) Two delay classes of messages.

    \caption{\small The illustrations show a DAG data structure, where messages are displayed as blocks, and child-parent relationships are displayed as arrows. Messages are sorted from left to right by their time of issuance. In \textbf{a)} all messages have a similar delay, while in \textbf{b)} a new class of message is introduced (marked in yellow) with noticeably increased delay, thus extending the typical time difference between parent and child. }
\label{fig:tangle}    
\end{minipage}\hfill
\end{figure}

Distributed  Ledger  Technologies  (DLTs) have gained much attention as a means to process and confirm data and transactions in a decentralised fashion. A fundamental component is the underlying data structure, which records messages either totally or partially ordered. Many DLTs, such as Bitcoin \cite{nakamoto2008bitcoin}, employ a blockchain structure, in which transactions are accumulated in blocks. These blocks are appended to each other creating a totally-ordered child-parent relationship, which makes the fate of the child block dependent on the parent. Certain DLTs such as IOTA, Nano, or Avalanche employ a more complex, partially-ordered Directed Acyclic Graph structure (DAG) \cite{Pervez2018}, \cite{Bencic2018}, \cite{boyen2016}, in which a message (or block) appends to several parent messages (or blocks), see Fig. \ref{fig:tangle}a).

Typically, it is assumed that messages require similar time for propagation, processing and creation, which can be summarized by a generic \textit{delay} \cite{popov2015}. 
While this may be reasonably precise when messages have the same content and similar size, this assumption may not hold in scenarios where  different types of messages exist or even when messages have similar content but different sizes ;
first, messages can have different roles and content ---such as utility messages, value transactions, or only generic data---, which can lead to differences in the processing time.
For example, a value transaction will have to pass additional checks compared to generic data, since a value transfer affects the token ownership in the ledger. 
Second, the propagation time may also depend on other factors, such as message size or prioritization of messages on the communication layer (as an example, in Bitcoin, the propagation time of a block depends on the byte size of the content
\cite{Decker2013}).
which Third, for rate control purposes, messages may have to be created with a proof of consumption of a scarce resource. Typically, the proofs of consumption used are Proof of Work (PoW) or Verifiable Delay Functions (VDFs) which, due to the computation time, add a delay. If several types of devices are present in the network, or messages require different difficulty levels for the PoW/VDF, the times for the message creation may differ significantly. 

Fig. \ref{fig:tangle}b) illustrates how the introduction of a new class of (yellow) messages with larger delay can extend the time until first reference, alter the DAG structure
and the size of the tip pool ---i.e., the pool of messages that were not referenced yet.
Specifically, the yellow class of messages
will be added to the tip pool
at a much later time compared to their issuance time, 
thus, effectively contributing
to the DAG much later than the original class of messages.
Moreover, the delay in the addition of this message \textit{to} the tip pool implies a delay in the removal of its parents \textit{from} the tip pool 
(messages are removed from the tip pool during inclusion of messages referencing them).
Thus, a second message with a smaller delay can also reference those same parents. Since this second message can be effectively added to the DAG before the old (and delayed) first message, the delayed message might not contribute to the tip pool removal at all. Generally, the larger the delay of a message
in comparison to other messages,
the smaller the contribution it has to the removal of tips from the tip pool.
These alterations in the DAG growth dynamics can lead to a significant change in some variables of interest, such as the number of tips or the width of the DAG. 
Since these attributes can have a major impact on the performance of the DLT, their accurate prediction and control are of vital importance.

\subsubsection*{Overview of the paper.}
In sections \ref{sec:discussion} and \ref{sec:quarantinetime}, we will focus on the first iteration of the IOTA 2.0 protocol \cite{2020coordicide} (a.k.a. Coordicide), and describe how it introduces two specific classes of messages with distinct delay times. Since the Coordicide is still an ongoing research project, progressing through several iterative development stages, it is worth noting that the substance of the delay classes may be significantly altered in subsequent protocol implementations. However, the implementation details described in this paper are aligned with the initially proposed protocol \cite{2020coordicide}.
Due to the generic dependency on the delay of messages, the derived results can be readily transferred to other scenarios and DLTs, where messages with differing delay times are present. 

In general, there can be several classes of messages with different delay times, which are introduced in Section \ref{sec:model}. We provide a generic model for $n$ different classes, that accurately predicts the tip pool size, and validate it against simulation and experimental results. Finally, in Section \ref{sec:control} we apply the model by describing a mechanism that enables the control of the tip pool size.

\section{Types of messages and their processing}\label{sec:discussion}

In the IOTA 2.0 protocol,\footnote{https://github.com/iotaledger/IOTA-2.0-Research-Specifications} two different classes of messages coexist: data messages and value messages. The first type consists of messages containing a payload with only data (called data payload), whereas the second one may also contain information about the transfer of funds and its associated unlock data, such as signatures. A payload containing this additional information is called a value payload.
Value messages are required to pass additional checks, which expose them to a voting mechanism if double-spends are attempted \cite{popov2019fpcbi,Capossele2021}. 
In this section, we briefly explain how these messages get handled and why transactions have to pass an additional voting filter. This will lead to Section \ref{sec:quarantinetime}, where the introduction of an additional delay time for value messages is explained.

\subsubsection*{Inclusion of messages into the ledger.}
For messages to be considered final, a mechanism to ensure the permanent addition to the ledger is required. 
If the majority of nodes processes the same (or similar) parts of the DAG, this requirement will be satisfied whenever a proportion $\theta$ of messages in the tip pool directly or indirectly references the message (since honest nodes would keep growing this same part of the DAG). We call this the inclusion criterion.

In this paper, we assume there are no restrictions for the addition of data messages to the ledger. Thus, data messages can be considered final once the inclusion criteria is satisfied. For value messages, the situation is more complicated, as we will see in the following section and Section \ref{sec:tipinclusion}.

\subsubsection*{Voting filter.} 

In IOTA 2.0, the transfer and allocation of funds are handled through an unspent transaction output (UTXO) model, which implies that an unspent output can only be spent once in the ledger, being considered thereafter as consumed. In an honest setting and without the existence of errors, we could apply the same as for data messages and assume that the inclusion criteria could be sufficient to consider transactions as final. However, due to user errors or in the presence of a malicious actor, it is possible that there are multiple transactions that attempt to spend the same output. These types of conflicting messages are called double-spends, and their occurrence requires the network to come to agreement on which transaction shall be accepted, or whether all should be rejected. This agreement is achieved by employing a voting protocol. Since the voting protocol effectively selects out certain transactions it forms a type of filter, i.e. the voting filter.
Once the voting filter validates a given value message, its inclusion will be guaranteed with high probability: since the nodes in the network have added the value message to their tip pool, the transaction should be eventually picked up by the tip selection algorithm and referenced by a majority of the network within a reasonable time. 

For the vast majority of situations, this is a sufficient finality criterion when dealing with double-spendings. However, it may be possible that a node may have an agreement failure on a transaction, thus, forming a wrong opinion about its inclusion into the ledger \cite{Capossele2021}. It is, therefore, reasonable to also require that the message inclusion criteria defined in the previous section should be satisfied.

In IOTA, the role of the voting filter is performed by the Fast Probabilistic Consensus (FPC) protocol \cite{popov2019fpcbi,Capossele2021}. On a more fundamental level, FPC constitutes the pre-consensus protocol, that allows nodes to come to an initial agreement on the state of a bit, by querying a random subset of nodes. With very high probability (see \cite{popov2019fpcbi}), after a finite number of rounds, the FPC will finalize on a boolean value, which is utilized for the decision on accepting or rejecting the transaction. 

A message that is rejected by the voting filter cannot enter the tip pool and must become orphaned, which means it should not be accounted for in the ledger. Note that any message that references the rejected transaction should not be included to the normal tip pool. Without any rescue mechanism for the referencing messages, such as re-attachments, these would also become orphaned. This orphanage issue can be  avoided to a large extent by introducing a quarantine time on value messages, which is discussed in the following section.

\section{Delay time for value messages}\label{sec:quarantinetime}

Messages are subject to several delay vectors before they can enter the tip pool of a node. First, the broadcast in the network does not happen instantly, but rather is dominated by a natural network delay that depends on several factors, such as geographical distance and distance in the P2P virtual network \cite{Mao2020}. Second, messages also need to be processed locally, are members in a waiting pool before being processed, or depend on messages that have not yet arrived. All of this adds together to a delay time, which results in an average delay time $h$.

In this section, we introduce and reason for an additional delay (or quarantine time) $d_Q$ for value messages, which effectively extends the delay time $h$ to $d=h+d_Q$. Specifically, in Section \ref{sec:sync}, we provide the setup of synchronicity assumptions that shape the requirements for the quarantine, whereas in Section \ref{sec:quarantineProcedure}, we describe some implementation details for the quarantine procedure.

\subsection{Synchronicity assumptions}\label{sec:sync}

For simplicity, we require in this paper that messages are delivered within a bound $d_Q/2$. Thus, in this paper, we do not treat the case where the network is partitioned for a time window larger than $d_Q/2$. We set $d_Q$ such that it is much larger than the average network delay and processing time of messages $h$. 
In more realistic conditions, this situation is more complicated, since packet losses may occur or faulty nodes might be present.
However, due to the interconnection and dependencies of messages, which are mapped by the DAG structure, nodes can identify missing messages and request them to their peers. 
This decreases the likelihood that a given node does not receive a message before $d_Q/2$.

For the dependencies on FPC, it is  sufficient for a super-majority of messages to arrive within $d_Q/2$ to maintain the integrity of the initial opinion. Moreover, for the FPC opinion exchange ---which is not part of the normal gossiping of messages---, the protocol will correctly finalize even if a part of those messages are lost. Thus, for the opinion exchange, it is sufficient to assume a probabilistic synchronous model, where we require that a large majority of opinion exchanges arrive within some bound time $t < d_Q/2$. The probabilistic character of this type of synchronicity assumption is given by the requirement that the large majority of communication occurs within this window, w.h.p. \cite{Capossele2021}. If the synchronicity assumption is not met (e.g. if a node did not receive a minimum required number of FPC responses), the node may ignore a voting round. In this sense, the security of the consensus protocol is ensured, while the liveness is temporarily halted.

\subsection{Quarantine procedure}\label{sec:quarantineProcedure}

The protocol component that is tasked with quarantining a transaction also oversees the initial opinion setting, and delegates whether the message is immediately added to the tip pool, if it first must pass the voting filter, or if it is rejected even before being processed by the voting filter. In theory one could also forward all transactions to the voting filter instead of employing a quarantine time. However, since the communication overhead for the voting protocol could become excessive, this option is not feasible if there are many transactions issued. Furthermore, even non-double spending transactions would have to pass the voting filter, which would impose an excessive delay time, i.e. much larger than $d_Q$.

\subsubsection*{Initial opinion setting.}\label{sec:likestatus}

The output that the voting filter provides on a given transaction is given in the form of a boolean value, \textit{liked} (for accepting the transaction) or \textit{disliked} (for rejecting the transaction). We can further decrease the probability for agreement failures by introducing an initial time window of length $d_Q/2$, during which a transaction's like-status is \textit{unknown}. The transaction is set to \textit{liked}, only if no other double-spending transaction arrives within $d_Q/2$. This ensures that, if there are double-spending transactions, at most one of them can be \textit{liked} by a large proportion of the nodes. We apply the following rules:
\begin{enumerate}
    \item A transaction is \textit{liked} after $d_Q/2$ if no other double-spending transaction arrives within $d_Q/2$, and \textit{disliked} otherwise.
    \item A transaction is \textit{disliked} on arrival if there is already a transaction that is spending the same output.
\end{enumerate}

\subsubsection*{Tip inclusion and voting filter activation.}\label{sec:tipinclusion}

We enforce a quarantine time $d_Q$ on an arriving transaction, during which the voting is not yet enabled and the transaction is not included to the tip pool yet. Once the quarantine time has elapsed, the transaction has to pass a tip inclusion check, which either will immediately pass the transaction to the tip pool or will initiate the voting filter instead. The outcome of the filter then determines whether the transaction is finally included into the tip pool. We apply the following rules:
\begin{enumerate}
    \item If no double spend transaction arrives within $d_Q$, the transaction passes a tip inclusion check and is added to the tip pool.
    \item If a double spend transaction arrives within $d_Q$, it fails the tip inclusion check, and it is only added to the tip pool if it is \textit{liked} by the voting filter.
\end{enumerate}
Note that the time between setting the \textit{like} status and performing the tip inclusion check is $d_Q/2$, which satisfies the requirements for the probabilistic synchronicity. Due to this window, a node that applies rule number 1) can be certain that the nodes that applied 2) ---instead of 1)---  will eventually \textit{like} the transaction, since it was already \textit{liked} by a super-majority of nodes that applied 1). Thus, FPC guarantees that the initially \textit{liked} transaction will become the winning transaction w.h.p.. The combined approach of rules 1) and 2) must be required, since value transactions must not enter the tip pool pre-maturely and get \textit{disliked} retrospectively. Contrarily, assume a transaction enters the tip pool and is approved by the node, but the transaction is later \textit{disliked}, then the transaction (as well as its referencing message) must be orphaned, i.e. fail to be included into the ledger.

\section{Delay classes and tip pool model}\label{sec:model}

In the previous sections we showed, using the example of the IOTA 2.0 protocol, how  the existence of different types of messages can result in differing delay times for them. In this section, we describe an analytical model of the tip pool size for the case where $n$ different classes are present. Furthermore, we investigate the case for $n=2$, which applies to the previous sections, in more detail.

The dynamics and message relationships within the DAG data structure affect several metrics that have an impact on the performance of the DLT. For example, the time and order in which messages arrive affect the number of references that they obtain. Generally speaking, nodes do not share a global view and, as a consequence, if a certain node considers a certain message as unreferenced, this may not be the case from the point of view of another node. The underlying cause of this is the existence of delays until a message is processed, which also differs from node to node. Furthermore, the tip pool size is also correlated to the time until first reference and, ultimately, the time until finalization of messages. In this paper, we will focus on the tip pool size.
We introduce \textit{delay classes} $C_i$, where the messages with this classification have a specific delay time $d_i$.
The introduction of delay classes creates a more complex scenario, where not always the first message to select a tip will be the message to remove this message from the tip pool. Since the delay introduced by the quarantine time of the value messages is order of magnitudes larger than the network and other processing-related delays, we can introduce ---in the case described in the previous section--- two delay classes $C_{\text{value}}$ and $C_{\text{data}}$, i.e. value  and data messages, with strongly distinct and constant delay times. This can lead to the situation that, if a fraction $p_{\text{value}}$ of the issued messages are value messages (which have an extended delay), the probability of a tip being removed from the tip pool by a value message can be significantly smaller than $p_{\text{value}}$. This affects the tip pool size and the time until the first reference of messages.

\subsection{General model for the tip pool size}
 Let $\mathcal{C}=\{C_i\}_{i=1,\dots,n}$ be the family of classes of messages. For each class $C_i$, we define the following variables:

\begin{itemize}
    \item $d_i$: delay of a message of class $C_i$ (w.l.g., we assume $d_i\leq d_{i+1}$)
    \item $k_i$: number of referenced messages,
    (i.e. parents), of each message of class $C_i$
    \item $p_i$: fraction of messages of class $C_i$ (thus, $\sum_{i=1}^n p_i=1$)
\end{itemize}

To issue a message, a node must select a certain number of messages $k_i$ as parents for it. The relation created through these parent-child references forms the DAG structure.
We now proceed to calculate the average tip pool size $L$. We model the different classes of messages' arrival as independent Poisson processes of rates $p_i\lambda$, for $i=1,\dots,n$. Thus, assuming that the size of the tip pool does not excessively deviate from the average value $L$, each new message of class $C_i$ will have a probability $k_i/L$ of referencing a given message in the tip pool (here, we implicitly assume that $L\gg k_i$, so that the probability of a tip being referenced by two or more references of the same incoming message is low). Thus, a tip sees the incoming references as independent Poisson processes of rates $\mu_i\coloneqq p_i\lambda k_i/L$ (for $i=1,\dots,n$). Generally, the number of references (i.e., children) a message receives can vary; however, since we are only interested in the tip status of a message, we are also only interested in the reference that removes the message from the tip pool. Now, let $S_1^i$ be the time until the first event of the Poisson process describing the reference arrival of class $C_i$ (from the point of view of a single tip). Then, the time $T$ until the removal of this tip from the tip pool will be given by the minimum between $d_i+S_1^i$ for $i=1,\dots,n$. Thus, we have 
\begin{equation}\label{eq.Ft}
    F_{T}(x) = 1-\prod_{i=1}^{n}\left[1-F_{d_i+S_1^i}(x)\right]
\end{equation}
\noindent where
\begin{equation}
        F_{d_i+S_1^i}(x) =\left[1-\exp(-\mu_i (x-d_i))\right]\mathds{1}_{(x>d_i)} \label{eq.ft2}
\end{equation}
\noindent By equations (\ref{eq.Ft}) and (\ref{eq.ft2}) and letting $a_i=\sum_{j=1}^{i}\mu_j$ and $b_i=\sum_{j=1}^{i}\mu_j d_j$:
\begin{align*}
    F_{T}(x) =& \sum_{i=1}^{n-1} \left[1-\exp\left(-a_i x+b_i) \right) \right]\mathds{1}_{(d_i<x\leq d_{i+1})}+  \left[1-\exp\left(-a_n x+b_n \right) \right]\mathds{1}_{(d_{n}<x)}
\end{align*}
\noindent Finally, the expected value of $T$ will be given by
\begin{align*}
    E(T)&=\int_{\mathbb{R}^+}x f_{T}(x) dx = d_{1}+\frac{1}{a_1}-\sum_{i=2}^{n} \exp\left(-d_ia_{i-1}+b_{i-1} \right)\left(\frac{1}{a_{i-1}}-\frac{1}{a_{i}}\right)
\end{align*}

\noindent By Little's Law, we have $L=E(T)\lambda$, implying the following implicit equation:
\begin{align}
    &L\left(1-\frac{1}{p_1 k_1}\right)= d_{1}\lambda-\sum_{i=2}^{n} e^{-\frac{\lambda}{L}\sum_{j=1}^{i-1} p_j k_j(d_i-d_j) }\left(\frac{L}{\sum_{j=1}^{i-1}p_j k_j}-\frac{L}{\sum_{j=1}^{i}p_j k_j}\right)
\end{align}

\subsection{Experimental and simulation validation}

In the case of the IOTA 2.0 protocol, a total of $n=2$ different delay classes with constant number of parents $k$ are observed. Applying the model developed in the last section to this specific case, we have the following implicit equation for the expected tip pool size $L$ (where $p\coloneqq p_{\text{value}}$):
\begin{align}\label{eq.implicit}
    &L= h\lambda+\frac{L }{k (1-p) }\left[1- p\exp\left(- \frac{(1-p)\lambda k d_Q}{L} \right)\right]
\end{align}

We proceed by finding the critical value $p^*$, that divides the domain in two regions $[0,p^*)$ and $[p^*,1]$ with different dominating classes of delays. We begin by linearising the solution of (\ref{eq.implicit}) around $p=0$ which gives us an approximate value of $L$ close to this point of
\begin{equation}\label{eqn:Lminus}
    L \approx \frac{\lambda h k}{k-1} + p \frac{\lambda h k}{ (k-1)^2}\left[1-\exp\left(- \frac{  d_Q (k-1)}{  h} \right)\right] \approx \frac{k}{k-1}\lambda h = L^-
\end{equation}

Analogously to the case above, if $p$ is large enough the tip pool size is increasingly controlled by value messages. As $p$ increases it becomes more likely that a given message is removed by a value message rather than a data message. More specifically, for large values of $p$, we have
\begin{align}
    L&\approx  L^+ = \frac{k\lambda }{k-1}(h+pd_Q)-\frac{k\lambda d_Q^2}{2(d_Q+h)}(1-p)\label{eqn:Lplus}
\end{align}

The intersection of the curves for equations (\ref{eqn:Lminus}) and (\ref{eqn:Lplus}) provides the approximate proportion of value messages $p^*$, for which the value messages start to have a noticeable impact on the tip pool size. Taking $L^-(p^*)=L^+(p^*)$, we have:
\begin{equation}\label{eqn:pcritical}
    p^*=\frac{d_Q(k-1)}{(2h+(k+1)d_Q)}
\end{equation}

 In order to validate the analytical derived equations above (specifically, \ref{eq.implicit}), we compare them against experimental as well as simulation results.
We run simulations for the Tangle for the 2-class message model, where one class has a fixed lower delay of $h$, and the second class has a fixed higher delay of $d$. Messages arrive randomly through a Poisson process. For each parameter set, we run the simulations for 1,000,000 message arrivals.

\begin{figure}
\vspace{-0.3cm}
\begin{minipage}{1.\textwidth}
\centering
\begin{subfigure}{.5\textwidth}
  \centering
  \includegraphics[width=\linewidth]{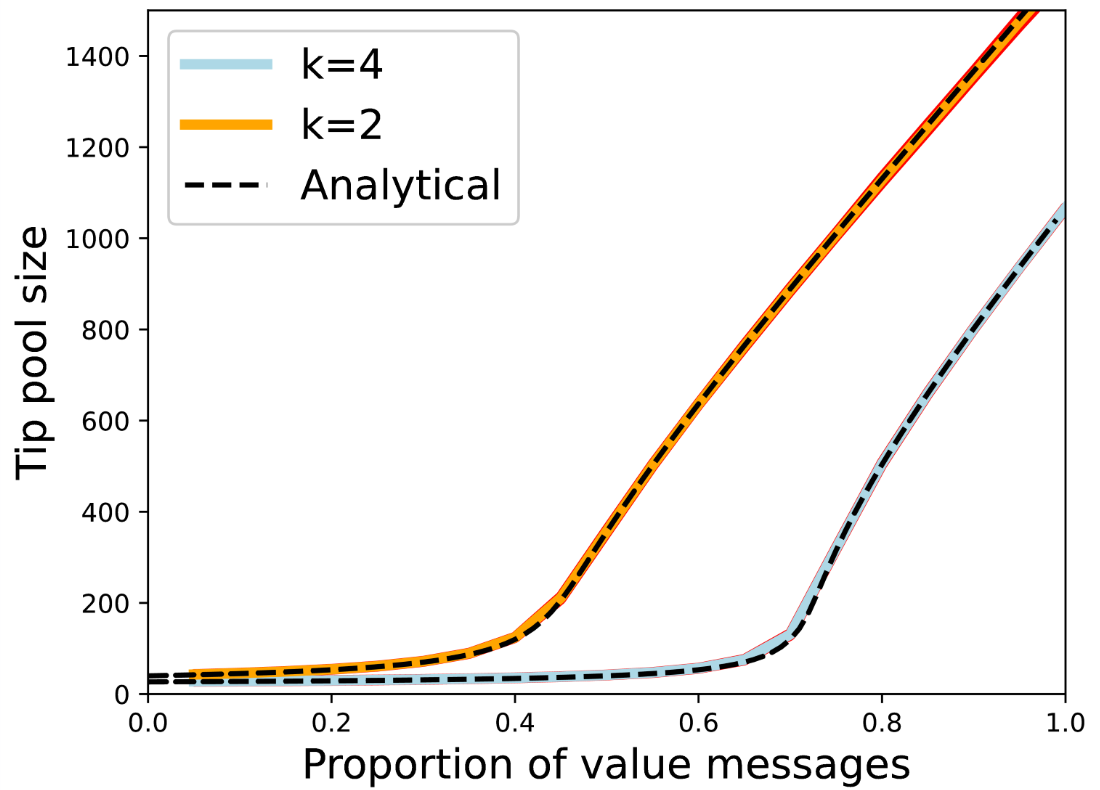}
  \caption{\small Simulation}
  \label{fig:modelsim}
\end{subfigure}%
\begin{subfigure}{.5\textwidth}
  \centering
  \includegraphics[width=\linewidth]{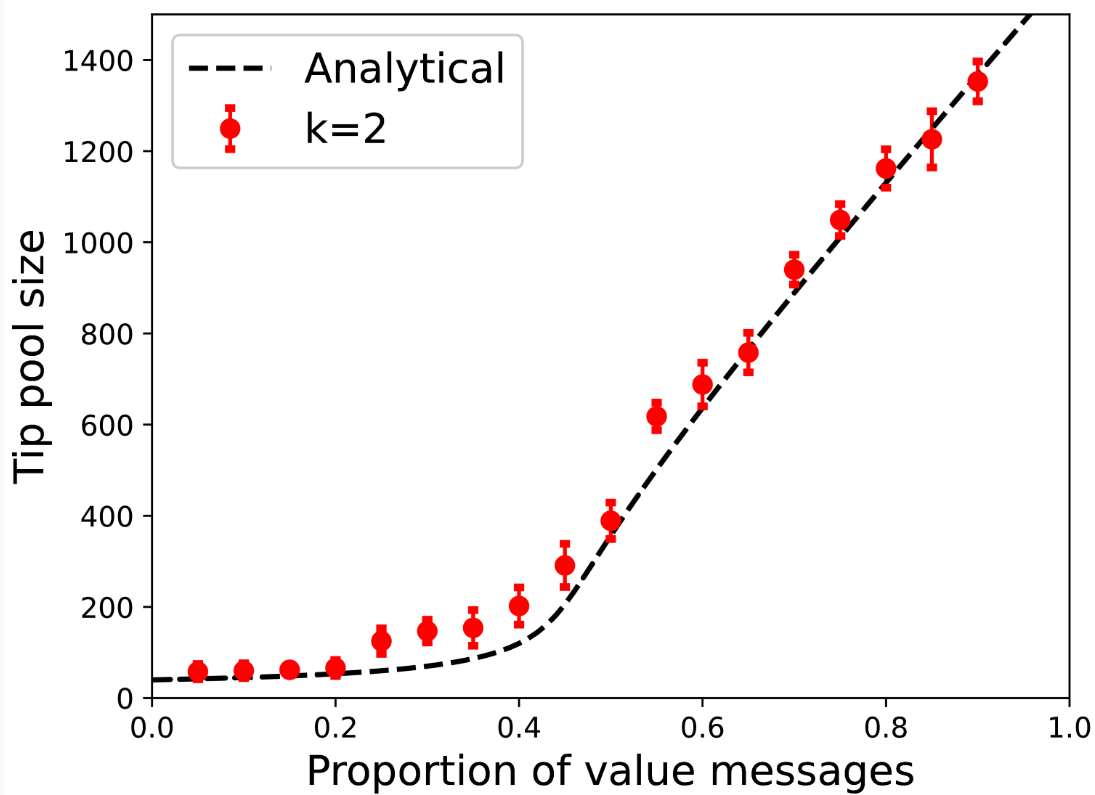}
  \caption{\small Experiment}
  \label{fig:modelexp}
\end{subfigure}
\caption{\small Tip pool size as a function of the proportion of value messages. We show data from  simulations (a), experiments (b), as well as analytical results for the parameters $\lambda=200Mps$, $h=0.1s$ and $d_Q=40h$. In figure a) we provide values for two different numbers of referenced messages $k$. The standard deviation is <3\% for the simulation. In figure b) results are marked along with their error for a confidence level of 95\%. }
\label{fig:model}
\end{minipage}\hfill
\end{figure}

Experimental values are obtained by measuring the tip pool of a node connected to a network that operates with the prototype implementation \textit{GoShimmer} of the IOTA 2.0 protocol\footnote{{https://github.com/iotaledger/goshimmer}}. \textit{GoShimmer} is a full node software, that is developed to test and validate the IOTA 2.0 solution for a fully decentralized DLT. For each parameter set, we issue messages at a combined rate for value and data of 200 messages per second (Mps).  We remove the data obtained for the ramp-up and down phase, which underestimates the tip pool size, since the data is not obtained from the stable phase, where the tip pool size stays constant.

Fig. \ref{fig:model} shows the tip pool size as a function of the proportion of value messages $p$, comparing the analytical, simulated (\ref{fig:modelsim}), and experimental (\ref{fig:modelexp}) results. The total rate of incoming messages $\lambda$ is measured in Mps. A good agreement between analytical prediction, experimental results, and simulation results can be observed. 
From the figure and by referring to equation (\ref{eq.implicit}) we can see that the tip pool size has a small and roughly constant slope close to $p=0$, i.e. the tip pools size is strongly dominated by the data messages in this region. Analogously, if $p$ is large enough the tip pool size is increasingly controlled by value messages. As $p$ increases it becomes more likely that a given message is removed by a value message rather than a data message. Furthermore, when $k$ is increased the tip pool can be kept stable for an increased proportion of value messages.  



\section{Controlling the tip pool size}\label{sec:control}

As we show in the previous section, the tip pool can vary significantly depending on the dominant class of messages. 
However, the tip pool size affects some important performance metrics in the DAG. For example, with an increased tip pool size, the width of the DAG increases, which, in turn, affects the time until the inclusion criterion is fulfilled. Therefore, it is desirable to keep the tip pool small.
This must be ensured even in the presence of fluctuations of incoming data and value messages, which can change which class of message is dominant. 

\begin{figure}[ht]
\vspace{-0.3cm}
\begin{minipage}{1.\textwidth}
\center
    \includegraphics[width=.67\textwidth,trim={0 0 0 0},clip]{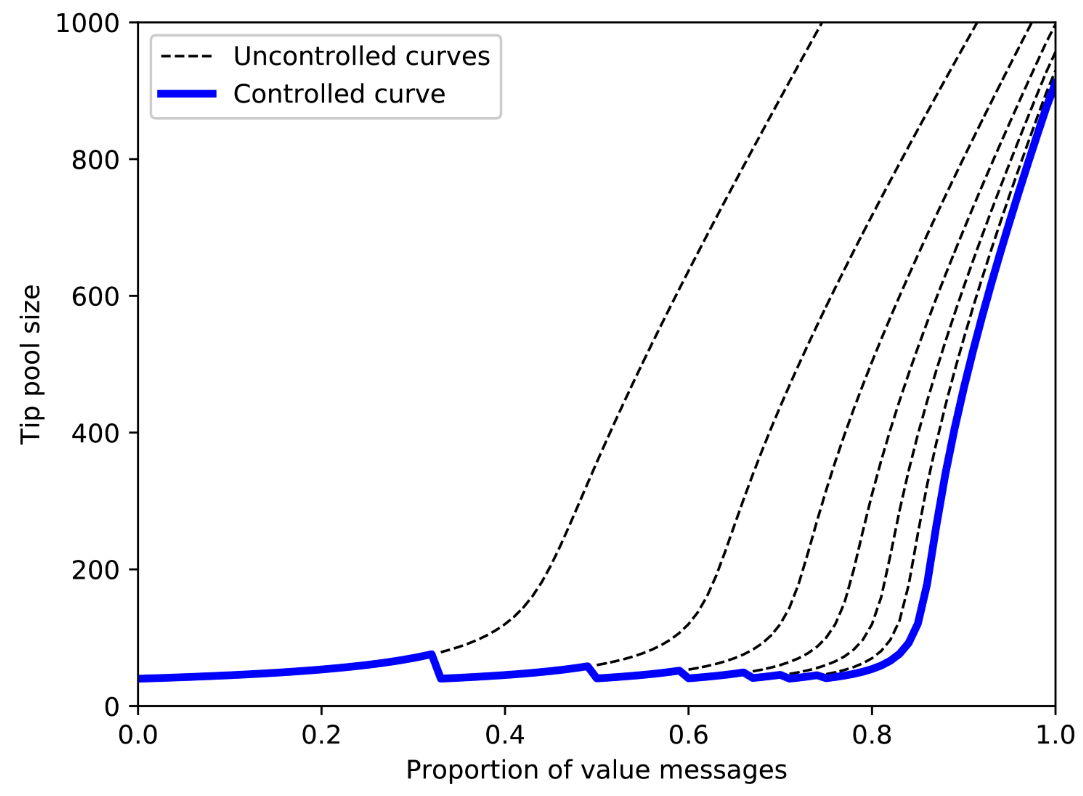}
    \caption{\small Tip pool size as a function of the proportion of value messages, when the number of parents is adapted and controlled (blue). The tip pool size is also indicated for the cases where the parent number is fixed (dashed). $\lambda=200Mps$, $h=0.1s$ and $d_Q=40h$. }
\label{fig:model2}    
\end{minipage}\hfill
\end{figure}

Applying this to the results of the previous section, we can observe that for a fixed number of parents $k$, $p$ can exceed the critical value $p^*$ occasionally. During such periods of excess of value messages, the tip pool size and the time until first reference could, therefore, increase substantially, if the parent number is fixed. 

From (\ref{eqn:pcritical}) we can see that for $h\ll d_Q$ the value for $p^*$ is independent of $h$, $d_Q$ and the absolute values of the transaction rates. 
By measuring a moving average of the value and data message rates, and calculating the moving average of the proportion of value messages $\Bar{p}$, a node can locally adapt $k$ ---up to a predefined maximum value $k_{max}$--- in its tip selection algorithm, such that $p^*(k)>\Bar{p}$. 
Algorithm \ref{alg:adaptiveParentNumber} shows the pseudo code that should be called before selecting the parents.
\begin{algorithm}[h]
\DontPrintSemicolon 
 Input: $\Bar{p}$\;
 k=2\;
 \While{ $p^*(k)$<$\Bar{p}$ AND k<$k_{max}$}  { k ++   }
 return k
\caption{Adaptive parent number control} \label{alg:adaptiveParentNumber}
\end{algorithm}

Note that increasing $k$ may increase the message size, since the parents have to explicitly be mentioned. Furthermore, increasing $k$ may also increase the processing time of the data messages and thus effectively increase $h$. However, this increase is negligible, since $h$ is typically dominated by other factors, such as the propagation time of the message through the network. If the majority of nodes in the network adopt this strategy, the tip pool sizes and the time for inclusion can be kept low in exchange for a minimally increased parent number. Fig. \ref{fig:model2} shows how the tip pool size would vary with the proportion of value messages, if the parent number can be variably controlled up to a maximum number of parents of $k_{max}=8$. For reference, the curves representing a fixed number of parents are also indicated. We show that using this procedure the tip pool can be kept stable at a small size for a much larger proportion of value messages.

\section{Conclusion}

Messages in DAG-based DLTs are appended to each other through parent-child relations. Certain aspects of the DAG structure created through this append process ---such as the width of the DAG and the pool of non-referenced messages---, depend on the delay of these messages. Since messages can have different purposes, such as transfer of value, utility, or generic data transfer, the processing time and delivery may create distinct classes in terms of perceived delay. 

In order to model the effects of the delay classes, we develop an analytical model for the generic case of $n$ distinct classes that can accurately predict certain metrics of the DAG structure, such as the tip pool size. We apply the model to the IOTA 2.0 protocol, in which value messages are treated differently compared to plain data messages. We show that simulation and experimental results agree well with the predicted values. For the simulation the standard deviation is <3\%, while for the experiment the error for a confidence level of 95\% is <3.3\% and <28.9\%  when the tip pool size is large and small, respectively. The shown validity of the model suggests that the model is suitable to study also other DAG scenarios, and where there is more than one message type present.

Through the analysis we show that the tip pool size depends on which class of message is dominant. The tip pool remains stable as long as a sufficient amount of the message class with a shorter delay is present. However, at some point, the tip pool may significantly increase due to the second class with a larger delay. Furthermore, a greater amount of messages with large delays can be supported, if the parent number is increased. This can be done statically, or dynamically through an adaptive mechanism.

\renewcommand{\abstractname}{\ackname}




\end{document}